\documentclass[sigconf,preprint]{acmart}
\AtBeginDocument{%
  }
    
\usepackage{xcolor} % For defining custom colors
\usepackage{listings}
\definecolor{codebg}{rgb}{0.95, 0.95, 0.95}
\definecolor{keywordcolor}{rgb}{0.13, 0.29, 0.53}
\lstdefinelanguage{Evaluation}{
    basicstyle=\ttfamily\footnotesize,
    keywordstyle=\bfseries,
    frame=single,
    breaklines=true,
    morekeywords={Conciseness, Numerical, Precision, Relevance, Factuality, Timeliness, Comprehensiveness, Clarity, Coherence, Insightfulness, Originality}
}
\usepackage{booktabs}  % 
\usepackage{graphicx}  % 
\usepackage{siunitx}   % 
\usepackage{hyperref} % Enables hyperlinking
\usepackage{cleveref}
\definecolor{kddblue}{RGB}{0,102,204} % A professional blue
\definecolor{kddgray}{RGB}{85,85,85}  % Neutral gray for less emphasized text

\hypersetup{
    colorlinks=true,
    linkcolor=kddblue, % For internal links and references
    citecolor=kddblue, % For citations
    urlcolor=kddgray   % For URLs
}
\crefname{section}{Sec.}{Sec.}
\crefname{appendix}{App.}{Apps.}
\crefname{theorem}{Thm.}{Thms.}
\crefname{proposition}{Prop.}{Props.}
\crefname{equation}{Eq.}{Eqs.}
\Crefname{equation}{Eq.}{Eqs.}
\crefname{table}{Tab.}{Tabs.}
\Crefname{table}{Tab.}{Tabs.}
\crefname{figure}{Fig.}{Figs.}
\Crefname{figure}{Fig.}{Figs.}
\crefname{algorithm}{Alg.}{Algs.}
\Crefname{algorithm}{Alg.}{Algs.}
\crefname{assumption}{Asm.}{Asms.}
\Crefname{assumption}{Asm.}{Asms.}
\crefname{mechanism}{Mech.}{Mechs.}
\Crefname{mechanism}{Mech.}{Mechs.}
\crefname{definition}{Def.}{Defs.}
\Crefname{definition}{Def.}{Defs.}
\newcommand{\ourmethod}{\texttt{Xinyu}}

\setcopyright{none}
\acmConference[]
\acmBooktitle{Preprint}

\acmISBN{}
\acmDOI{}

\begin{document}

%%
%% The "title" command has an optional parameter,
%% allowing the author to define a "short title" to be used in page headers.
\title{Xinyu AI Search: Enhanced Relevance and Comprehensive Results with Rich Answer Presentations}

%%
%% The "author" command and its associated commands are used to define
%% the authors and their affiliations.
%% Of note is the shared affiliation of the first two authors, and the
%% "authornote" and "authornotemark" commands
%% used to denote shared contribution to the research.
\author{Bo Tang}
\authornote{Co-first author.}
\email{tangbo@mail.ustc.edu.cn}
\affiliation{%
  \institution{AIDS and SIAR, University of Science and Technology of China}
  \city{Suzhou}
  \country{China}
}

\author{Junyi Zhu}
\authornotemark[1]
\email{junyizhu.ai@gmail.com}
\affiliation{%
  \institution{ESAT-PSI, KU Leuven}
  \city{Leuven}
  \country{Belgium}
}

\author{Chenyang Xi, Yunhang Ge}
% \email{xicy@iaar.ac.cn}
% \author{Yunhang Ge}
% \email{geyh@iaar.ac.cn}
\email{firstname.lastname@iaar.ac.cn}
\affiliation{%
  \institution{Institute for Advanced Algorithms Research}
  \city{Shanghai}
  \country{China}
}

\author{Jiahao Wu}
\email{jiahao.wu@connect.polyu.hk}
\affiliation{%
  \institution{The Hong Kong Polytechnic University}
  \city{Hong Kong}
  \country{China}
}

\author{Yuchen Feng, Yijun Niu}
\author{Wenqiang Wei, Yu Yu}
\author{Chunyu Li, Zehao Lin}
\email{firstname.lastname@iaar.ac.cn}
\affiliation{%
  \institution{Institute for Advanced Algorithms Research}
  \city{Shanghai}
  \country{China}
}

\author{Hao Wu, Ning Liao}

\author{Yebin Yang, Jiajia Wang}

\author{Zhiyu Li, Feiyu Xiong}
% \author{Zhiyu Li}
% \email{lizy@iaar.ac.cn}
% \author{Feiyu Xiong}

\email{firstname.lastname@iaar.ac.cn}
\affiliation{%
  \institution{Institute for Advanced Algorithms Research}
  \city{Shanghai}
  \country{China}
}

\author{Jingrun Chen}
\email{jingrunchen@ustc.edu.cn}
\authornote{Corresponding author.}
\affiliation{%
  \institution{AIDS and SIAR, University of Science and Technology of China}
  \city{Suzhou}
  \country{China}
}

\renewcommand{\shortauthors}{Bo Tang and Junyi Zhu et al.}

%%
%% The abstract is a short summary of the work to be presented in the
%% article.
\begin{abstract}
Traditional search engines struggle to synthesize fragmented information for complex queries, while generative AI search engines face challenges in relevance, comprehensiveness, and presentation. To address these limitations, we introduce Xinyu AI Search, a novel system that incorporates a query-decomposition graph to dynamically break down complex queries into sub-queries, enabling stepwise retrieval and generation. Our retrieval pipeline enhances diversity through multi-source aggregation and query expansion, while filtering and re-ranking strategies optimize passage relevance. Additionally, Xinyu AI Search introduces a novel approach for fine-grained, precise built-in citation and innovates in result presentation by integrating timeline visualization and textual-visual choreography. Evaluated on recent real-world queries, Xinyu AI Search outperforms eight existing technologies in human assessments, excelling in relevance, comprehensiveness, and insightfulness. Ablation studies validate the necessity of its key sub-modules. Our work presents the first comprehensive framework for generative AI search engines, bridging retrieval, generation, and user-centric presentation.
\end{abstract}

\begin{teaserfigure}
    \centering 
  \includegraphics[width=1\textwidth]{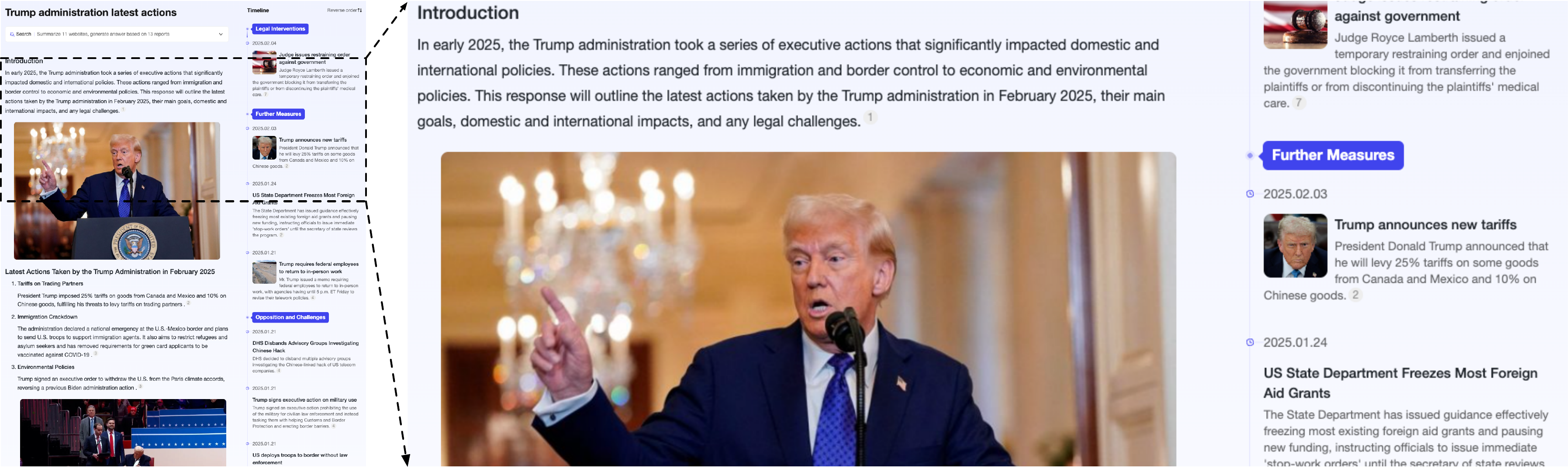}
  \caption{Online evaluation (on February 5th) of the Xinyu AI search system for the query ‘Trump administration latest actions.’ \ourmethod{} features built-in citation, timeline visualization (right column), and a textual-visual choreography mechanism.}
  \label{fig:teaser}
\end{teaserfigure}

\maketitle

\section{Introduction}
In the era of information, a significant volume of events, knowledge, and resources has been digitized and made accessible through the Internet. As users continually strive to quickly and accurately locate relevant information within this rapidly growing digital ecosystem, the development of search engines has emerged as a critical solution to meet their fundamental need for efficient and reliable information retrieval~\cite{google,Manning_Raghavan_Schütze_2008}.
However, traditional search engines often face challenges in addressing complex or ambiguous queries. 
% They struggle to process unstructured data effectively and lack the ability to capture semantic and contextual relationships. 
Furthermore, these systems typically present results as a ranked list, requiring users to manually synthesize information from diverse sources. This significantly increases comprehension efforts, particularly in scenarios where aggregating fragmented information from multiple resources is required.

Previously, the field of language modeling has been significantly advanced by the development of autoregressive models based on Transformer architectures \cite{vaswani2017attention,Radford2018ImprovingLU}. These architectures enable efficient parallel processing of sequences and facilitate large-scale unsupervised pretraining~\cite{devlin2018bert,Radford2018ImprovingLU,radford2019language,kaplan2020scaling}, leading to the emergence of intelligent behaviors. Additionally, reinforcement learning from human feedback \cite{christiano2017deep,ouyang2022training} has further refined these models by aligning their outputs with human preferences and values. Together, these advancements have empowered machines to comprehend long-form text, interpret human intent, and generate responses that closely resemble human communication.
Modern large language models (LLMs) have demonstrated human-level performance in tasks such as reading comprehension and reasoning within specific contexts~\cite{devlin2018bert,liu2019roberta,yang2019xlnet,street2024llms}. Their vast parameters also enable the encoding of extensive knowledge~\cite{brown2020language,T5}. Despite these strengths, LLMs continue to face critical challenges, including outdated knowledge and the generation of hallucinated content~\cite{lewis2020retrieval,zhang2023hallucination}. These issues significantly undermine their reliability and limit their effectiveness in real-world applications.

More recently, retrieval-augmented generation (RAG) has emerged as a promising framework that integrates information retrieval techniques, such as search engines, with LLMs~\cite{lewis2020retrieval}. Studies have demonstrated that with externally retrieved non-parametric information, RAG can substantially reduce hallucinations in generated outputs while enabling LLMs to provide up-to-date information through in-context learning~\cite{jiang-etal-2023-active,ayala-bechard-2024-reducing,ni-etal-2024-llms}. In turn, LLMs can enhance search quality by rewriting queries to better align with search engine requirements, improve readability of search results by synthesizing fragmented information from multiple retrieved sources, and summarize long-form texts~\cite{ma2023query,yang-etal-2018-hotpotqa,devlin2018bert,brown2020language}. 

\subsection{The Development and Challenges of Generative AI Search Engines}
Building on the concept of RAG, generative AI search engines such as Perplexity AI~\cite{perplexity}, Tiangong AI~\cite{tiangong}, and Metaso~\cite{metaso} have emerged to provide synthesized answers using LLMs, rather than merely returning links like traditional search engines. While conversational LLM-based products such as ChatGPT~\cite{chatgpt} also employ RAG to access up-to-date information and enhance factual accuracy, generative AI search engines distinguish themselves by prioritizing comprehensiveness and improving the overall reading experience through advancements in answer presentation. For instance, incorporating built-in citations that link directly to sources to build user trust in search results. A survey conducted in the United States found that over a quarter of adults considered switching to AI search engines in 2023~\cite{statista2023}.

Although RAG and LLMs complement each other, the diversity of user queries makes it challenging to generate satisfactory responses solely based on retrieved documents. Moreover, commercial AI search engines may produce inaccurate or unfaithful answer. To analyze these limitations, we collected 300 queries across eight domains and identified several common issues in existing technologies, such as Perplexity AI. The results are presented in \cref{fig:issues}.

\subsection{Our Approach to Generative AI Search}
\begin{figure}
  \includegraphics[width=\linewidth]{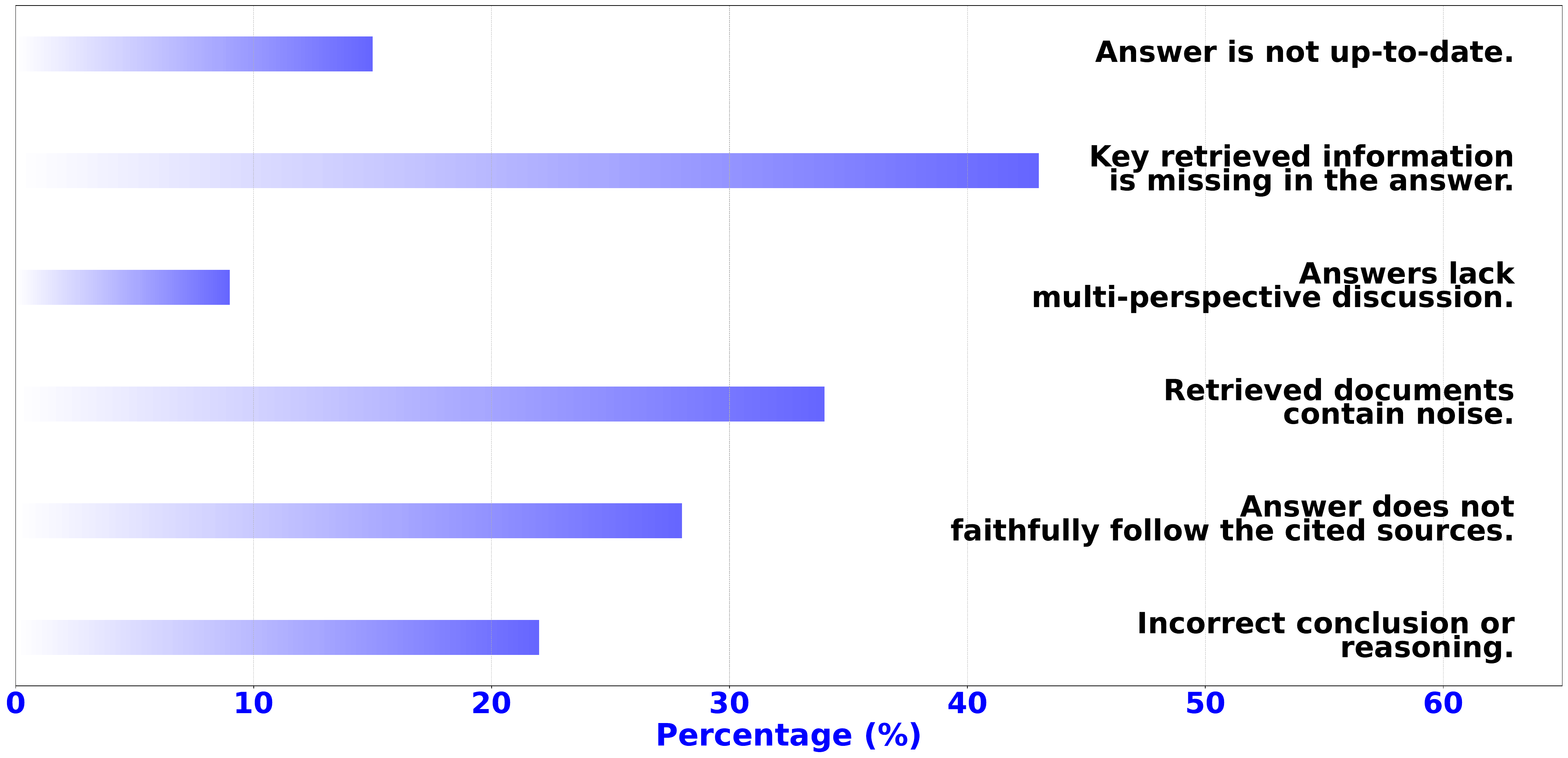}
Figure  \caption{Common issues in generative AI search answers. }
  \label{fig:issues}
  \vspace{-0.5em}
\end{figure}
The issues outlined in \cref{fig:issues} affect the relevance and comprehensiveness of answers to queries, along with several other factors that degrade text generation quality. Beyond these challenges, improving the reading experience through enhanced presentation of generated results represents another key area for advancement in generative AI search. To address these issues, we develop \ourmethod{} (which indicates “a new way to present” in Chinese). In this paper, we provide a detailed breakdown of our proposed method and demonstrate how we orchestrate its workflow. An online test showcase is presented in \cref{fig:teaser}.

\paragraph{\textbf{Our contributions can be summarized as follows:}}
\textbf{(1)} We systematically decompose this domain into specific subproblems and provide detailed descriptions of our solutions, including prompt design, data preparation, and model training, to facilitate future research and applications.
\textbf{(2)} We introduce novel approaches, including query decomposition graphs, timeline visualization, textual-visual choreography, and built-in citations, to address multiple core challenges.
\textbf{(3)} Experimental results demonstrate that \ourmethod{} is competitive with existing technologies, and we conduct extensive ablation studies to evaluate the effectiveness of individual components.
\textbf{(4)} To the best of our knowledge, this is the first paper to provide a comprehensive disclosure of a generative AI search engine system.\footnote{While online blog articles offer illustrative overviews of AI search systems (e.g., Perplexity AI), and Lepton AI's released code serves primarily as a demonstrative example, our work is distinct. We provide a full-stack technical disclosure and comprehensive evaluations, offering greater value for reproducibility and as a research reference.}

\subsubsection{\textbf{Paper Organization}} Main terms are clarified below. In \cref{sec:rw}, we discuss related work. In \cref{sec:method}, we present our method. In \cref{sec:exp} we compare our method with existing technologies and conduct ablation studies. Finally, we conclude in \cref{sec:conclusion}.

\subsubsection{\textbf{Terminology}}
We define the key terms and concepts below:

\noindent \textcolor{keywordcolor}{Query}: The user’s input query.

\noindent \textcolor{keywordcolor}{Sub-query}: A query decomposed from the user’s input query.

\noindent \textcolor{keywordcolor}{Retrieval query}: A query submitted to the search interface.

\noindent \textcolor{keywordcolor}{Retrieved document}: Content accessed via links returned by the search engine.

\noindent \textcolor{keywordcolor}{Retrieved passage}: A segment of text in the retrieved document.

\begin{figure*}
\centering
\includegraphics[width=\textwidth]{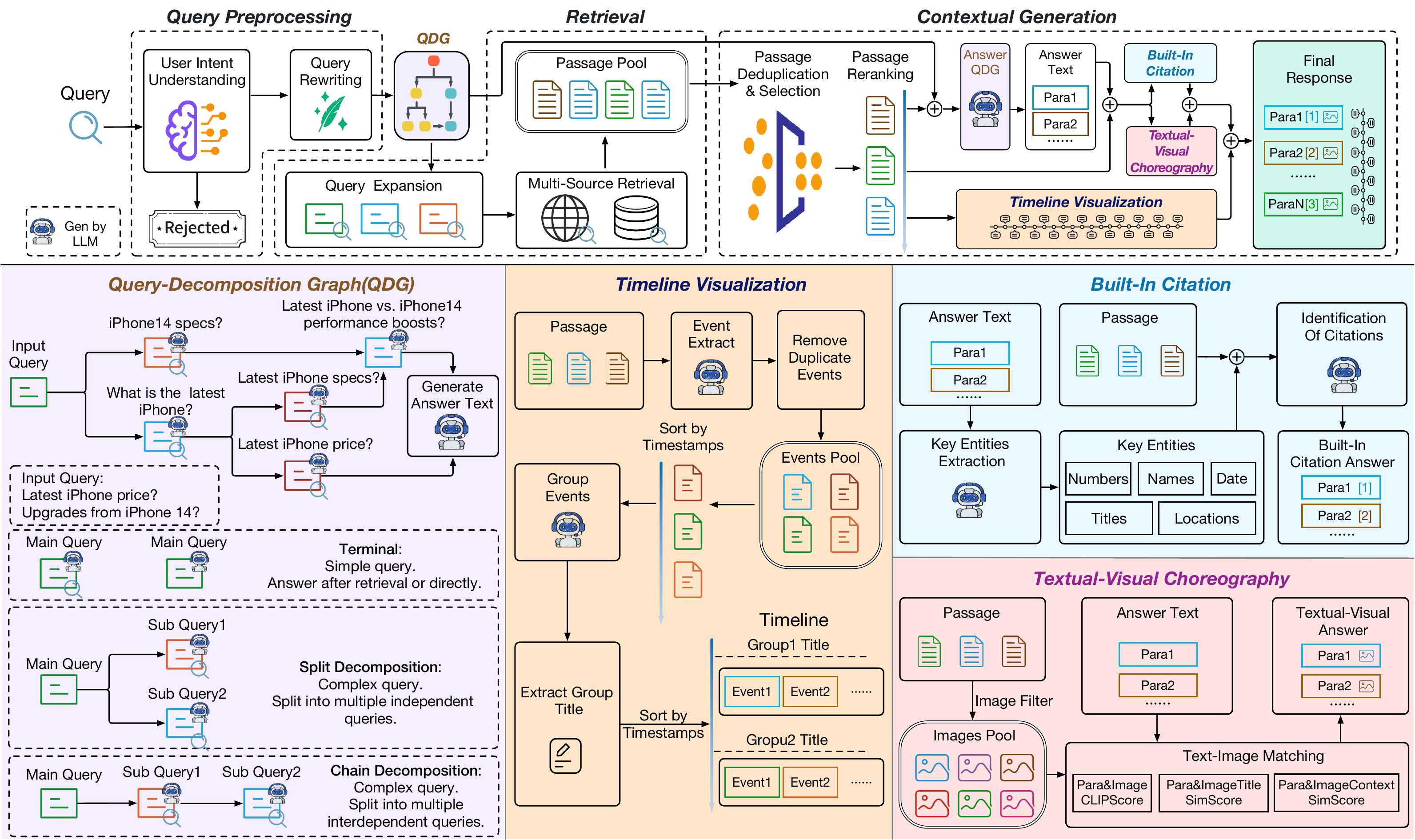}
  \caption{\ourmethod{} AI search framework. The upper row illustrates the full response pipeline, while the lower row provides a more detailed depiction of several novel approaches integrated into this framework.}
  \label{fig:framework}
\end{figure*}
\section{Related Work}
\label{sec:rw}
Constructing an generative AI search engine involves the design of three main components: retrieval, contextual generation and orchestration. Since these components relate to a broad range of research topics. We briefly introduce the work closely related with this paper. A more detailed survey can be found at \cite{gao2024retrievalaugmented}.

\subsection{Retrieval}
\label{sec:rw:retrieval}
Effective retrieval is critical to system performance, as the quality of retrieved in formation significantly shapes the final output. Query rewriting techniques aim to transform user queries into more precise and retrieval-friendly formats, addressing ambiguities and enhancing alignment with indexed data~\cite{gao-etal-2023-precise,ma-etal-2023-query,peng2024large,zheng2024take}. Similarly, query expansion enriches the input by generating alternative or supplementary queries, ensuring the retrieval of a broader and more contextually relevant set of documents~\cite{dhuliawala-etal-2024-chain,wang-etal-2023-query2doc,jagerman2023query}. 
The choice of retrieval sources also impacts system performance, with strategies leveraging unstructured data, semi-structured data, and structured knowledge graphs to provide domain-specific and fine-grained knowledge~\cite{wang2023knowledgpt,zha2023tablegpt,luo2023augmented,he2024gretriever}. Lastly, when database construction is needed, effective chunking strategies, metadata enrichment, and hierarchical indexing are considered to ensure that retrieval components operate efficiently~\cite{theja2023chunk,microsoft2024rag,ibm2024metadata}.

\subsection{Contextual Generation}
\label{sec:rw:gen}
After retrieving documents for a query, the generation process relies on their context to produce responses that are accurate and well-informed. Studies show that irrelevant information in the references can distract the model and lead to inaccurate answers~\cite{yoran2024making,cuconasu2024power}. Additionally, LLMs allocate varying levels  of attention to different sections of the prompt, making the placement of relevant information crucial to the quality of the generated response~\cite{liu-etal-2024-lost}. Reference filtering techniques aim to eliminate irrelevant or noisy retrieved documents, ensuring only pertinent information is considered for generation~\cite{cui2023chatlaw,ma2023large}. Context selection focuses on identifying the most relevant portions of the retrieved context while discarding less pertinent parts, thereby optimizing the model’s input~\cite{yang-etal-2023-prca,jiang-etal-2024-longllmlingua}. Reference reranking further reorganizes the retrieved information to position the most relevant content prominently, improving the quality of responses~\cite{zhuang-etal-2023-open,gao2023chat}. Some work employ fine-tuning to adapt a language model for specific tasks or domains, or to align its outputs with desired formats and styles, thereby achieving superior task performance~\cite{du-ji-2022-retrieval,li-etal-2023-structure,zhu-etal-2024-fastmem}.

\subsection{Orchestration}
Simple pipelines that directly generate responses from retrieved results often fall short, prompting research into auxiliary components and the orchestration of more sophisticated workflows. Iterative workflows involve alternating between retrieval and generation processes, progressively enriching the context by utilizing generated text or intermediate results to refine subsequent retrievals~\cite{shao-etal-2023-enhancing}. Adaptive workflows enhance system flexibility by dynamically determining the necessity of retrieval based on the context of the query, often incorporating mechanisms for self-assessment and adjustment~\cite{jiang-etal-2023-active,asai2024selfrag,nakano2021webgpt}. Recursive workflows break down complex queries into smaller, interdependent subtasks, iteratively resolving each to produce comprehensive and logically structured responses~\cite{trivedi-etal-2023-interleaving,kim2023tree}. Specifically, the chain-of-knowledge strategy first generates rationales for answering a query, then leverages retrieval results to refine these rationales and deduce the final response~\cite{li2024chainofknowledge}.

\section{Technical Approach of \ourmethod{} AI Search}
\label{sec:method}
Next, we elaborate on our method. In \cref{sec:preproc}, we present the query preprocessing steps. In \cref{sec:tree}, we describe how the workflow is orchestrated using our proposed query-decomposition graph. Our retrieval system is detailed in \cref{sec:retrieval}, followed by an explanation of the steps taken to enhance generation quality using the retrieved documents in \cref{sec:generation}. Finally, we discuss specific components designed for rich answer presentation in \cref{sec:reading}.

In \ourmethod{}, instruction fine-tuned LLMs, embedding models, and rerankers are extensively utilized across various tasks. We leverage multiple open-source models of different sizes, balancing performance and efficiency. Task-specific fine-tuning can further enhance model effectiveness. However, given the complexity of the overall system, fine-tuning all models individually is prohibitively expensive. To maintain cost efficiency, we adopt a unified data preparation and model fine-tuning framework for several key tasks. Below, we introduce this framework, with further details provided in the appendices, which are referenced in subsequent sections.

\paragraph{Unified Framework of Data Preparation and Model Fine-tuning}
Let $\mathcal{D}$ denote the training dataset. Data preparation involves collecting existing public datasets, as well as generating synthetic data (or labeling data) using stronger models (e.g., larger models) and subsequently refining data quality through expert selection~\cite{albalak2024a}. For fine-tuning generative LLMs, the training data consists of input and ground-truth answer pairs $(x, y) \in \mathcal{D}$. Given model parameters $\theta$, we optimize them using the next-token prediction (NTP) objective:

\begin{equation}
\label{eq:ntp}
\mathcal{L}_{\text{NTP}}(\theta) = -\mathbb{E}_{(x,y) \sim \mathcal{D}} \left[ \sum_{t=1}^{|y|} \log p_{\theta}(y_t | x, y_{<t}) \right],
\end{equation}
where $ y_{<t} $ represents preceding tokens. For fine-tuning reranker models, the training data consists of an anchor sample $x$, a positive sample $x^+$, and $N$ negatives $x^-_{1:N}$, i.e., $(x, x^+, x^-_{1:N}) \in \mathcal{D}$. Let $h_{\theta}(\cdot)$ denote the score function. We optimize the network to correctly predict the positive sample among the negatives using a cross-entropy loss over the scored candidates:

\begin{equation}
\label{eq:reranker}
\mathcal{L}_{\text{Re}}(\theta) = -\mathbb{E}_{(x, x^+, x^-_{1:N}) \sim \mathcal{D}} 
\left[ \log \frac{e^{h_{\theta}(x, x^+)}}{e^{h_{\theta}(x, x^+)} + \sum_{i=1}^{N} e^{h_{\theta}(x, x_i^-)}} \right].
\end{equation}

% InfoNCE contrastive loss~\cite{oord2018representation} with temperature $\tau$ and similarity metric $s(\cdot, \cdot)$:

% \begin{equation}
%     \label{eq:contrast}
%     \mathcal{L}_{\text{Cont}}(\theta) = -\mathbb{E}_{(x,x^+,x^-_{1:N}) \sim \mathcal{D}} \left[ \log \frac{e^{s(h_{\theta}(x), h_{\theta}(x^+)) / \tau}}{\sum_{x'\in \{x^+, x^-_{1:N}\}} e^{s(h_{\theta}(x), h_{\theta}(x')) / \tau}}\right].
% \end{equation}
% where $B$ contains both positive sample $x^+$ and negatives. 
% For fine-tuning reranker models, training data is in form of query $x$, document $y$ and score $s$ triplet, i.e.\ $(x, y, s)\in\mathcal{D}$. We train the model to predict the score:

\subsection{Query preprocessing}
\label{sec:preproc}
When a query is input by the user, initial steps are conducted to ensure that the query is safe and harmless. Additionally, query disambiguation can help improve the search quality when query is passed to the search module.

\subsubsection{User Intent Understanding}
As illustrated in the top row of \cref{fig:framework}, understanding user intent is the first step after a query is submitted. This module initially filters out harmful or unsafe queries, such as those that violate legal or ethical standards or compromise privacy. Additionally, if a query is ambiguous or lacks specificity, the system prompts the user with clarifying questions or options to refine their intent. For example, if a user submits the query “The current state of the economy,” the system suggests options to specify a region of interest (see~\cref{fig:disambiguation}).
To support these functionalities, we fine-tune a generative LLM (Qwen2.5-14B) to analyze queries, make judgments, suggest potential clarifying options, and output the results in JSON format. Query rejection and refinement are based on parsing relevant keywords. A detailed discussion of the fine-tuning process for this task is provided in \cref{app:ft:intent}.

\subsubsection{Query Rewriting}
After user intent understanding, we conduct query rewriting to align the query with the search engine requirement. To address the issue of geo-temporal queries, e.g.\ "Shanghai news from last week",  we supplement meta data about the user local time and location, then instruct a generative LLM (Qwen2.5-14B)  to specify these information in the query if needed.

% \subsubsection{Temporal Scope Identification}
% We find that determining the temporal scope of retrieved documents is crucial for generating accurate and temporally relevant responses. Incorporating the user’s local time through query rewriting can help address some of these challenges. Building on this step, we also retrieve documents based on the query and identify their most relevant temporal scope. For example, when a user submits the query, “What policies has Trump introduced during his new term?”, we leverage a generative language model to analyze the retrieved documents and extract the temporal scope (e.g., 20/01/2025 to the present) in JSON format. Rather than using the temporal scope to further refine the query—an approach that does not consistently improve retrieval quality—we pass it to the reference consistency check process during the generation stage, which will ensure noisy documents falling outside the identified temporal scope are filtered out (c.f. \cref{sec:consistency}). For the details of the instruction prompt and model fine-tuning, please refer to.

\subsection{Query-Decomposition Graph (QDG)}
\label{sec:tree}
To overcome the limitations of naive RAG in capturing nuanced and multi-faceted information for complex questions, we propose a novel and practical decomposition strategy that breaks down the user’s query into sub-queries and answers them step by step. We refer to this approach as the query-decomposition graph (QDG).

As shown in \cref{fig:framework}, after query rewriting, a single query is transformed into a QDG. A specific example is provided at the bottom left of \cref{fig:framework}. In the QDG, nodes represent sub-queries, while directed edges indicate dependencies.  Given a query, we use a fine-tuned generative LLM (Qwen2.5-72B) to construct the corresponding QDG by defining nodes and their pairwise relationships. The LLM is instructed with the QDG definition and few-shot examples. Guided by the prompt, the LLM applies the following decomposition strategies: \textbf{1)} \textit{Chain decomposition:} A query is sequentially decomposed into a series of sub-queries, where each parent node provides preliminary information for its child node and the descendant nodes, e.g.\ least-to-most~\cite{zhou2023leasttomost}. \textbf{2)} \textit{Split decomposition:} The query is divided into multiple independent sub-queries. \textbf{3)} \textit{Terminal:} The input query is elementary, and decomposition is unnecessary. Further details on the instruction prompts and model fine-tuning are provided in \cref{app:qdt,app:ft:qdg}. We find that the success rate of generating valid QDGs approaches 1, but we also perform validation and reiterate the generation process if the check fails.

The QDG defines the workflow for subsequent generation processes: the parent node is processed first, while the child node generates its output based on retrieved documents, ancestor sub-queries, and their corresponding answers. With QDG as its core, \ourmethod{} facilitates hierarchical reasoning and evidence aggregation, ensuring a comprehensive and logically consistent resolution of complex queries.

\subsection{Retrieval}
\label{sec:retrieval}
After constructing the QDG, we aggregate all sub-queries and perform retrieval. To ensure the retrieved documents capture diverse details necessary for generating answers, we enhance retrieval diversity through query expansion and multi-source retrieval, as discussed below. For an illustration, refer to the top row of \cref{fig:framework}.

\subsubsection{Query Expansion}
\label{sec:expansion}
We first enrich the sub-queries by asking an LLM to generate multiple retrieval queries revolving around a given sub-query. Specifically, the LLM is instructed to act as a subject matter expert in an university, expanding the given query to create related questions that assess students’ comprehensive understanding of the topic across multiple dimensions: 1) content mastery, 2) understanding of key elements, 3) contextual analysis, and 4) extended thinking. 
% Details of the instruction prompt are provided in \cref{app:expansion}.

\subsubsection{Multi-source retrieval}
\label{sec:multi-source}
To ensure the real-time relevance of retrieved information, we invoke search engine APIs. Since search engines employ varied ranking algorithms, we submit each retrieval query to multiple sources simultaneously to obtain more comprehensive content. Directly feeding raw web page content into an LLM can lead to high perplexity and degraded response quality. To make the input content LLM-friendly and safe, we implement a robust content filtering pipeline. This process involves removing disruptive elements, filtering sensitive or extraneous information, and standardizing formatting. More details on the filtering rules are provided in \cref{app:filter_rule}.
After filtering, we segment documents into passages using the \verb|RecursiveCharacterTextSplitter| method from \verb|LangChain|\cite{Chase_LangChain_2022}. We adopt a small chunk size of 350 with a relatively large 25\% overlap to optimize the performance of the text embedding model, following the study by Azure AI Search~\cite{Berntson2023}.

% \textcolor{red} {https://cloud.tencent.com/developer/article/1449213}

\subsection{Contextual Generation}
\label{sec:generation}
After retrieval, the retrieved passages are assigned to their corresponding sub-queries in the QDG. A fine-tuned LLM then generates responses for each sub-query following the dependency structure, ensuring that parent nodes are processed before their child nodes.
Notably, retrieved passages vary in relevance and often contain duplicate information, which can distract the model. Moreover, LLMs allocate different levels of attention to various sections of the input~\cite{liu-etal-2024-lost}. To mitigate these issues, we implement passage deduplication, selection, and re-ranking, as detailed below. For an illustration, refer to the top row of \cref{fig:framework}.

\subsubsection{Passage Deduplication}
\label{sec:dedup}
Different retrieved documents may exhibit content homogenization, such as sharing the same viewpoints or reproducing information from a common source. To mitigate redundancy across passages, we first perform deduplication. Specifically, we use a fine-tuned text embedding model (bge-large-zh~\cite{bge_embedding}) to compute embeddings for the passages and then calculate pairwise cosine similarities. Our objective is to identify the largest subset of passages in which no two passages have a similarity score exceeding 0.8. Finding the optimal solution to this problem corresponds to solving the maximum independent set problem, which is NP-hard. To improve computational efficiency, we adopt a greedy algorithm that processes each passage sequentially, retaining it only if its similarity to all previously retained passages remains below 0.8. Details of fine-tuning are provided in \cref{app:fine-tune-rerank}.

\subsubsection{Passage Selection}
\label{sec:selection}
Since passages may contain irrelevant information, we mitigate noise by ranking their relevance to the sub-query. Specifically, we compute a weighted average of keyword frequency and TF-IDF scores, where keywords (e.g., time and location) are extracted from the sub-query using an LLM (Qwen2.5-14B). At this stage, we retain the top 70\% of the most relevant passages.

\subsubsection{Passage Rerank}
\label{sec:rerank}
Since LLMs allocate more attention to information at the edges of a prompt and tend to lose focus in the middle~\cite{liu-etal-2024-lost}, we further refine the retrieved passages presented to the LLM by sorting them based on their similarity to the sub-query. This ranking is performed using a fine-tuned reranker model (bge-reranker-v2-m3~\cite{chen2024bge,li2023making}). Details on dataset preparation and fine-tuning are provided in \cref{app:fine-tune-rerank}.

\subsubsection{Answer Generation}
After passage re-ranking, passages are appended to their respective sub-queries, and responses are generated in the order dictated by the QDG using a fine-tuned generative LLM (Qwen2.5-72B). If a sub-query has a parent node, the Q\&A results of all ancestor nodes are inserted before the retrieved passages. An example is provided in \cref{app:qdt}. Once all terminal nodes complete generation, their questions and answers are concatenated, appended to the main query, and used to generate the final response. Details on the fine-tuning process are provided in \cref{app:ft:generation}.

\subsection{Rich Answer Presentations}
\label{sec:reading}
Traditional chatbots often rely on linear text stacking, which can impose a high cognitive load on users. Given that AI-powered search engines facilitate extensive knowledge transmission, integrating cognitive scaffolding is essential to support user comprehension. Cognitive science research has shown that structured information and multimodal presentations enhance the efficiency of information assimilation~\cite{Prieto2018,JohnCognitive,Mayer2014}. Moreover, since mitigating hallucinations in LLMs remains challenging~\cite{zhang2023hallucination}, aiding users in result verification and fostering confidence in synthesized outputs is crucial. To address these challenges, we incorporate timeline visualizations, textual-visual choreography, and built-in citations, as discussed below, to optimize the reading experience.
% Notably, the timeline visualization and textual-visual choreography are distinctive features that set our approach apart from existing technologies.

\subsubsection{Built-In Citation}
\label{sec:reading:cite}
% Synthesized responses often summarize information from multiple documents. Built-in citations help users access additional details and verify the accuracy of the synthesized content.
A straightforward approach to citation generation involves instructing the LLM to produce citations on the fly, as implemented by Lepton AI~\cite{LeptonSearch2025}. However, our initial evaluation indicates that this method exhibits a high error rate. Furthermore, we observe that some existing systems, such as Perplexity AI, place citations at the end of a paragraph, potentially detaching references from the corresponding evidence. To enhance both citation accuracy and granularity, we propose a novel citation scheme.

As illustrated in the bottom right of \cref{fig:framework}, \ourmethod{} decouples answer generation from citation attachment. Our pipeline employs two models. The first model, an SLM (Qwen2.5-3B), extracts key entities (e.g., dates, locations, names) from the generated answer on a \textit{sentence-by-sentence basis}. If a sentence contains extractable entities, a second SLM (Qwen2.5-3B) identifies citations based on these entities, its orignal sentence, and retrieved documents. Both models have been fine-tuned for their respective tasks. Details of prompt and fine-tuning are provided in \cref{app:promp:cite,app:ft:cite}. In cases where no entities are extracted from a sentence, we adopt a fallback method that computes the sentence embedding using bge-large-zh and assigns a citation if its cosine similarity with a retrieved document exceeds 0.6.
To reduce latency, we implement an asynchronous processing strategy that runs citation assignment in parallel with answer generation (albeit with a one-sentence delay).

\subsubsection{Timeline Visualization}
\label{sec:reading:timeline}
In online search scenarios focused on news and events, 
integrating timeline visualizations enables users to better understand the evolution and context of events. We propose a novel timeline visualization scheme as illustrated in the bottom middle of \cref{fig:framework}. First, we collect all retrieved passages following the passage selection (see \cref{sec:selection}). Next, we instruct an LLM (Qwen2.5-14B) to extract any event time mentioned in each passage and to generate a corresponding title and summary. If a passage does not explicitly mention a time, we resort to using the document’s report time as extracted by the same LLM. Passages lacking temporal information in both the passage and the document are discarded. Because the retained passages may describe the same content, we then employ \texttt{bge-large-zh} to compute text embedding of the concatenated title and summary and calculate pairwise cosine similarities across passages. Passages with a similarity score exceeding 0.9 are merged by discarding the one with the later timestamp, resulting in a list of distinct events with timestamps. To make timeline visualization more structured, we further instruct the LLM to group these events and derive relevant keywords based on their summaries. Finally, we present the event titles for each group, sorted according to their timestamps.

\begin{table}[t!]
    \centering
    \caption{Pearson correlation coefficients between human and LLM scores for different evaluation criteria.}
    \label{tab:pearson_correlation}
    \vspace{-0.3em}
    \resizebox{0.8\linewidth}{!}{
    \begin{tabular}{l c l c}
        \specialrule{1.2pt}{0pt}{0pt} % Thicker top rule
        \textbf{Metric} & \textbf{Value} & \textbf{Metric} & \textbf{Value} \\
        \midrule
        Comprehensiveness         & 0.679 &
        Conciseness              & 0.787 \\
        Numerical Precision     & 0.741  & Clarity              & 0.737  \\
        Relevance               & 0.807  & Coherence            & 0.746  \\
        Factuality      & 0.831  & Insightfulness       & 0.610  \\
        Timeliness             & 0.759  &                      &        \\
        \specialrule{1.2pt}{0pt}{0pt} % Thicker bottom rule
    \end{tabular}
    }
\end{table}
% \begin{figure}[t!]
% \centering
% \includegraphics[width=0.9\linewidth]{figures/classification.pdf}
%   \caption{Domain distribution of 300 test queries.}
%   \label{fig:domain_dist}
% \end{figure}
\begin{figure}[t!]
\centering
% \vspace{-0.2em}
\includegraphics[width=0.9\linewidth]{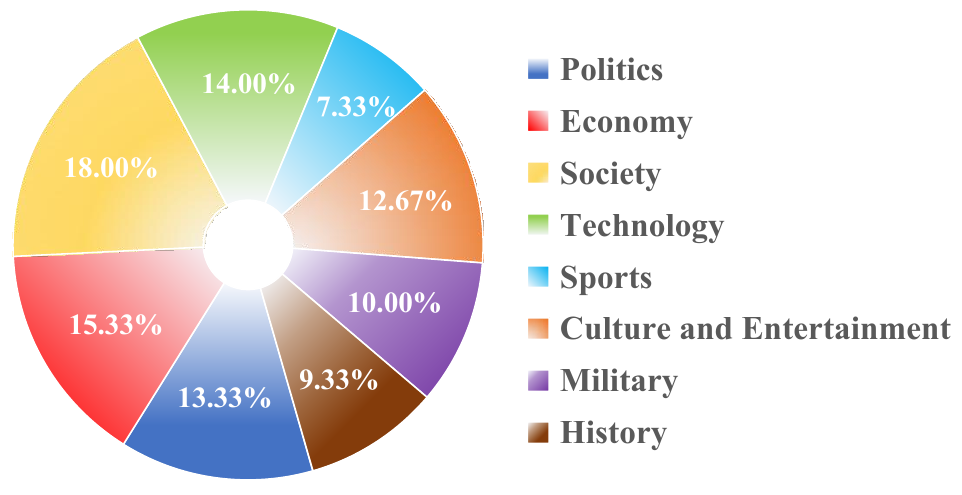}
  \caption{Domain distribution of 300 test queries.}
  \label{fig:domain_dist}
  % \vspace{-0.2em}
\end{figure}
\subsubsection{Textual-Visual Choreography}
\label{sec:reading:multimodal}
A picture is worth a thousand words. As illustrated in the bottom right of \cref{fig:framework}, \ourmethod{} integrates relevant images into textual responses to enhance information assimilation. These images are extracted from retrieved documents. To ensure quality and relevance, we first filter out noisy images, retaining only those of high quality. Specifically, a rule-based filtering algorithm eliminates logos, icons, and low-resolution images. Subsequently, we compute the similarity of the textual description associated with the image with the main query using bge-reranker-v2-m3 and remove those smaller than 0.3. To determine the optimal placement of images, we compute the pairwise similarity between generated answer paragraphs and candidate images. This computation involves a weighted average of three measures: (1) the embedding cosine similarity between the generated text paragraph and the image, computed using the chinese-clip-vit-huge-patch14; (2) the estimated similarity between the synthesized paragraph and the retrieved document’s title, obtained via bge-reranker-v2-m3; and (3) the embedding cosine similarity between the synthesized paragraph and the retrieved document’s text using bge-large-zh. Pairwise similarities are assembled into a matrix. Then we determine the optimal image-to-text alignment using Hungarian algorithm~\cite{hugarian}.

\begin{table*}[t!]
    \centering
    \caption{Multi-faceted comparison of different approaches. Higher value indicates better performance, 10 is the maximum.}
    \label{tab:performance_comparison}
    \resizebox{\textwidth}{!}{
    \begin{tabular}{l c c c c c c c c c c}
        \specialrule{1.1pt}{0pt}{0pt} % Thicker bottom rule
        \small \textbf{Model} & \small \textbf{Conciseness} & \small \textbf{Numerical Precision} & \small \textbf{Relevance} & \small \textbf{Factuality} & \small \textbf{Timeliness} & \small \textbf{Comprehensiveness} & \small \textbf{Clarity} & \small \textbf{Coherence} & \small \textbf{Insightfulness} & \small \textbf{Average} \\
        \midrule
        \small{Perplexity AI}~\cite{perplexity} &  \textbf{9.851} & 9.630 & 9.436 & 8.524 & 8.553 & 7.284 & 9.612 & 9.853 & 6.543 & 8.810 \\
        \small{Tiangong AI~\cite{tiangong}} & 9.840 &  \textbf{9.722} & 7.812 & 8.924 & 8.103 & 8.020 & 9.604 & 9.802 & 6.535 & 8.707  \\
        \small Ernie Bot~\cite{yiyan} & 9.770 & 9.320 & 8.883 & 8.028 & 8.406 & 7.798 & 9.524 & \textbf{9.900} & 5.963 & 8.621  \\
        \small KIMI~\cite{kimi} & 9.840 & 9.515 & 8.529 & 8.224 & 8.966 & 8.155 & 9.223 & 9.709 & 6.796 & 8.773\\
        \small Metaso~\cite{metaso} & 9.760 & 8.941 & 8.515 & 7.408 & 8.403 & 5.689 & 9.383 & 9.689 & 4.759 & 8.061 \\
        \small{ChatGLM~\cite{chatglm}} & 9.810 & 9.420 & 8.949 & \textbf{9.124} & 8.346 & 6.168 & 9.533 & 9.726 & 5.047 & 8.458 \\
        \small{Baichuan~\cite{baichuan}} & 9.660 & 9.596 & 6.486 & 7.612 & 8.220 & 8.252 & 9.223 & 9.612 & 6.117 & 8.309 \\
        \small{Tongyi~\cite{tongyi}} &9.803 & 9.009 & 7.586 & 7.212 & 8.194 & 7.677 & 9.293 & 9.899 & 5.859 & 8.281 \\

        \small{Xinyu (Ours)} & 9.813 & 9.714 & \textbf{9.533} & 8.932 & \textbf{9.205} & \textbf{9.143} & \textbf{9.633} & 9.810 & \textbf{7.333} & \textbf{9.235} \\
        \specialrule{1.2pt}{0pt}{0pt} 
    \end{tabular}
    }
\end{table*}
\begin{table*}[t!]
    \centering
    \begin{minipage}{0.70\textwidth} % Wider table
    \vspace{0pt plus 1fill}
    \caption{Comparison of timeline visualization. Except wall time, higher value is better.}
    \label{tab:event_based_comparison}
    \resizebox{\linewidth}{!}{
    \begin{tabular}{l c c c c c c }
        \specialrule{1.2pt}{0pt}{0pt} % Thicker top rule
        \small \textbf{Approach} & 
        \small \textbf{Timeliness} & 
        \small \textbf{Comprehensiveness} & 
        \small \textbf{Clarity} & 
        \small \textbf{Event Count} &
        \small \textbf{Precision} & 
        \small \textbf{Wall Time (s) $\downarrow$} \\
        \midrule
        \small CHRONOS~\cite{wu2025unfolding}       & 7.02  & 5.79  & 6.00  & 5.12  & 79\%  & 67.44  \\
        \small{Xinyu (Ours)}  & \textbf{8.07} & \textbf{8.08}  & \textbf{8.24}  & \textbf{10.14} & \textbf{84}\% & \textbf{33.27}   \\
        \specialrule{1.2pt}{0pt}{0pt} % Thicker bottom rule
    \end{tabular}
    }
    \end{minipage}
    \hfill % Horizontal spacing
    \begin{minipage}{0.29\textwidth}
    \vspace{0pt plus 1fill}% Narrower table
    \caption{Comparison of textual-visual choreography.}
    \label{tab:image_inclusion_comparison}
    \resizebox{\linewidth}{!}{ % Make the table full width
    \begin{tabular}{l c c}
        \specialrule{1pt}{0pt}{0pt} % Thicker top rule
        \small \textbf{Approach} & 
        \small \textbf{Inclusion (\%) $\uparrow$} & 
        \small \textbf{Precision (\%) $\uparrow$} \\
        \midrule
        Metaso~\cite{metaso}      & 3.0  & 72.2  \\
        \small{Xinyu (Ours)}        & \textbf{80.0} & \textbf{90.0}  \\
        \specialrule{1pt}{0pt}{0pt} % Thicker bottom rule
    \end{tabular}
    }
    \end{minipage}
\end{table*}
\begin{table}[t!]
    \centering
    \caption{Comparison of built-in citation.}
    \label{tab:citation_comparison}
    \resizebox{0.7\linewidth}{!}{ % Adjust to full width
    \begin{tabular}{l c c}
        \specialrule{1.2pt}{0pt}{0pt} % Thicker top rule
        \small \textbf{Model} & 
        \small \textbf{ Density (\%) $\uparrow$} & 
        \small \textbf{ Precision (\%)} $\uparrow$ \\
        \midrule
        \small Perplexity AI~\cite{perplexity}  & 46.6  & 82.1  \\
        \small Metaso~\cite{metaso}             & 59.5  & 49.7 \\
        \small Tiangong~\cite{tiangong}         & 27.0  & 90.8  \\
        \small Baichuan~\cite{baichuan}         & 45.7  &\textbf{90.9}  \\
        \small KIMI~\cite{kimi}                 & 41.4 & 72.9  \\
        \small Xinyu (Ours)                     & \textbf{67.2}  & 90.4  \\
        \specialrule{1.2pt}{0pt}{0pt} % Thicker bottom rule
    \end{tabular}
    } % Close resizebox
\end{table}
\begin{table*}[t]
    \centering
    \caption{Ablation study of sub-models in our approach, "$-$" indicates skipping the sub-module.}
    \label{tab:ablation_study}
    \resizebox{\linewidth}{!}{ % Make the table full width
    \begin{tabular}{l c c c c c c c c c c}
        \specialrule{1.2pt}{0pt}{0pt} % Thicker top rule
        \small \textbf{Variant} & 
        \small \textbf{Conciseness} & 
        \small \textbf{Numerical Precision} & 
        \small \textbf{Relevance} & 
        \small \textbf{Factuality} & 
        \small \textbf{Timeliness} & 
        \small \textbf{Comprehensiveness} & 
        \small \textbf{Clarity} & 
        \small \textbf{Coherence} & 
        \small \textbf{Insightfulness} & 
        \small \textbf{Average} \\
        \midrule
        \small Full Approach       & 9.880 & 9.547 & 9.547 & 9.731 & 8.300 & 8.533 & 9.900 & 9.747 & 7.107 &  {9.143} \\
        \small {$-$ Query Preprocessing} & 9.810 &9.422 &9.497 & 9.646 &8.279 &8.423 & 9.891 & 9.637& 6.993& 9.066\\
        \small {$-$ Query Expansion}   & 9.793 & 9.300 &  {9.593} &  {9.626} & 8.300 & 8.493 & 9.867 & 9.780 & 6.827 & 9.064 \\
        \small $-$ QDG               & 9.780 &  {9.607} & 9.413 & 9.731 &  {8.320} & 8.620 & 9.860 &  {9.827} & 6.993 & 9.127 \\
        \small $-$ Passage Selection & 9.833 & 9.473 & 9.513 & 9.717 & 8.207 & 8.613 & 9.847 & 9.787 & 7.060 & 9.118 \\
        \small $-$ Passage Rerank    & 9.827 & 9.587 & 9.587 & 9.731 & 8.220 &  {8.587} & 9.873 & 9.800 & 6.987 & 9.132 \\
        \specialrule{1.2pt}{0pt}{0pt} % Thicker bottom rule
    \end{tabular}
    } % Close resizebox
\end{table*}

\begin{table*}[t]
    \centering
    \caption{Ablation study of replacing our fine-tuned LLMs with proprietary models. The best results for each metric are bolded.}
    \label{tab:ablation_model}
    \resizebox{\linewidth}{!}{ % Make the table full width
    \begin{tabular}{l c c c c c c c c c c}
        \specialrule{1.2pt}{0pt}{0pt} % Thicker top rule
        \small \textbf{Model} & 
        \small \textbf{Conciseness} & 
        \small \textbf{Numerical Precision} & 
        \small \textbf{Relevance} & 
        \small \textbf{Factuality} & 
        \small \textbf{Timeliness} & 
        \small \textbf{Comprehensiveness} & 
        \small \textbf{Clarity} & 
        \small \textbf{Coherence} & 
        \small \textbf{Insightfulness} & 
        \small \textbf{Average} \\
        \midrule
        \small GPT-4O~\cite{chatgpt}           & 9.828 & 9.425 & \textbf{9.621} & 9.433 & 7.973 & 8.473 & 9.753 & 9.717 & 6.520 & 8.972 \\
        \small{Qwen 2.5-72B~\cite{yang2024qwen2}}     & 9.780 & 9.290 & 9.463 & 8.987 & 8.053 & 8.140 & 9.893 & 9.633 & 6.687 & 8.881 \\
        \small{Xinyu (Ours)}     & \textbf{9.880} & \textbf{9.547} & 9.547 & \textbf{9.731} & \textbf{8.300} & \textbf{8.533} & \textbf{9.900} & \textbf{9.747} & \textbf{7.107} & \textbf{9.142} \\
        \specialrule{1.2pt}{0pt}{0pt} % Thicker bottom rule
    \end{tabular}
    } % Close resizebox
\end{table*}

\section{Online Deployment and Experiments}
\label{sec:exp}
\paragraph{Online Deployment} \ourmethod{} AI search engine was initially developed for Chinese users. We have since launched an English version by converting Chinese prompts into English. This approach has shown surprisingly good performance, likely due to the multilingual capabilities of modern generative and embedding models. However, we believe that fine-tuning language-specific models can further enhance the system, and this optimization is currently in progress.

\Cref{fig:teaser} illustrates \ourmethod{}’s interface with an online test case.

\paragraph{Test Cases}
We collected a set of test queries covering eight domains. Since many user queries during deployment are related to trending news topics, we also gathered a set of recent queries for evaluation to compare product performance under real-time conditions. The query domain distribution is shown in \cref{fig:domain_dist}. As our expert evaluators are native Chinese speakers, the numerical evaluation results are based on Chinese queries. 

\paragraph{Multi-faceted Evaluation Criteria}
Because it is difficult to establish a gold-standard answer for a generative AI search engine, we adopted rating criteria for evaluating the generated answers rather than computing a match to a fixed answer. We invited experts with journalism backgrounds and master’s degrees to develop rating criteria that reflect a multi-faceted evaluation of the generated answers, including: \textbf{(1)} Conciseness, \textbf{(2)} Numerical Precision, \textbf{(3)} Relevance, \textbf{(4)} Factuality, \textbf{(5)} Timeliness, \textbf{(6)} Comprehensiveness, \textbf{(7)} Clarity, \textbf{(8)} Coherence, and \textbf{(9)} Insightfulness. More detailed definitions for each dimension are provided in the \cref{app:eval}.

\paragraph{LLM Evaluation}
Evaluating the generated answers for all experiments using multi-faceted criteria by human experts is prohibitively expensive. Therefore, we employed an LLM to evaluate the generated text for a subset of experiments. LLM evaluation has been adopted in many studies~\cite{gu2024survey}. We used GPT-4O (gpt-4-0125-preview) to assess the results by instructing it to provide point-by-point reasoning and computing its final score. We set temperature to 0, more details about the prompt are provided in the \cref{app:ft:eval}. \Cref{tab:pearson_correlation} shows that the LLM evaluation is highly correlated with human evaluation based on the results of \cref{tab:performance_comparison,tab:performance_comparison_llm}. 

\paragraph{Baselines}
We compare our approach with eight existing technologies, including: \textbf{Generative AI search engines}: Perplexity AI~\cite{perplexity}, Metaso~\cite{metaso}, Tiangong AI~\cite{tiangong}, ChatGLM~\cite{chatglm}, and Tongyi~\cite{tongyi}; and \textbf{Conversational LLMs with RAG}: KIMI~\cite{kimi}, Ernie Bot~\cite{yiyan}, and Baichuan~\cite{baichuan}. For Perplexity AI, we selected GPT-4O~\cite{chatgpt} as the backend model.

% \subsection{Interface Demonstration}

\subsection{Comparison with Existing Technologies}

\subsubsection{Multi-Faceted Evaluation of the Generated Answer.}
We compare \ourmethod{} with existing technologies by inviting human experts to evaluate answers generated by different approaches based on the multi-faceted evaluation criteria. The results, presented in \Cref{tab:performance_comparison}, show that \ourmethod{} performs competitively while achieving the highest average score (\textbf{9.235} vs.\ 8.810). Notably, \ourmethod{} significantly outperforms other methods in  comprehensiveness (\textbf{9.143} vs.\ 8.252), and insightfulness (\textbf{7.333} vs.\ 6.796). Additionally, we conduct an LLM-based evaluation using GPT-4O, where the model assesses the generated text according to the same rating criteria. The results are reported in \Cref{tab:performance_comparison_llm}.

\subsubsection{Representation Enhancement}
\paragraph{Built-In Citation}
We evaluate our approach using two key metrics. The first, citation precision, measures whether the provided evidence is genuinely supported by the cited source. The second, citation density, quantifies the proportion of sentences containing citations relative to the total number of sentences. Citation density reflects two factors: (1) the extent to which the generated answer relies on retrieved information and (2) the placement of citations.
Some existing systems, such as Perplexity AI, often position citations at the end of a paragraph, making it difficult for users to trace specific claims, especially when multiple citations correspond to different parts of a paragraph stack. In such cases, citation density is also lower. As shown in \cref{tab:citation_comparison}, \ourmethod{} achieves significantly higher citation density (\textbf{67.2} vs. 59.5) while maintaining competitive citation precision.

\paragraph{Timeline Visualization}
A recent study introduces CHRONOS for timeline generation~\cite{wu2025unfolding}. We compare our method against CHRONOS using three multi-faceted evaluation criteria—timeliness, comprehensiveness, and clarity—assessed by human evaluators. Additionally, we evaluate event count to measure the system’s ability to identify multiple events, precision to assess the relevance of extracted events to the query, and wall time of online deployments to gauge computational efficiency.
\Cref{tab:event_based_comparison} demonstrates that \ourmethod{} significantly outperforms CHRONOS across multiple dimensions.

\paragraph{Textual-Visual Choreography}
Among the baselines, Metaso~\cite{metaso} also implements textual-visual choreography. We compare our method against it by evaluating two metrics: inclusion (the rate at which images are incorporated into the generated answers) and precision (the percentage of included images that are contextually relevant). As shown in \cref{tab:image_inclusion_comparison}, \ourmethod{} outperforms Metaso significantly on both metrics.

\subsection{Ablation Study}
\begin{figure}
\centering
\includegraphics[width=0.7\linewidth]{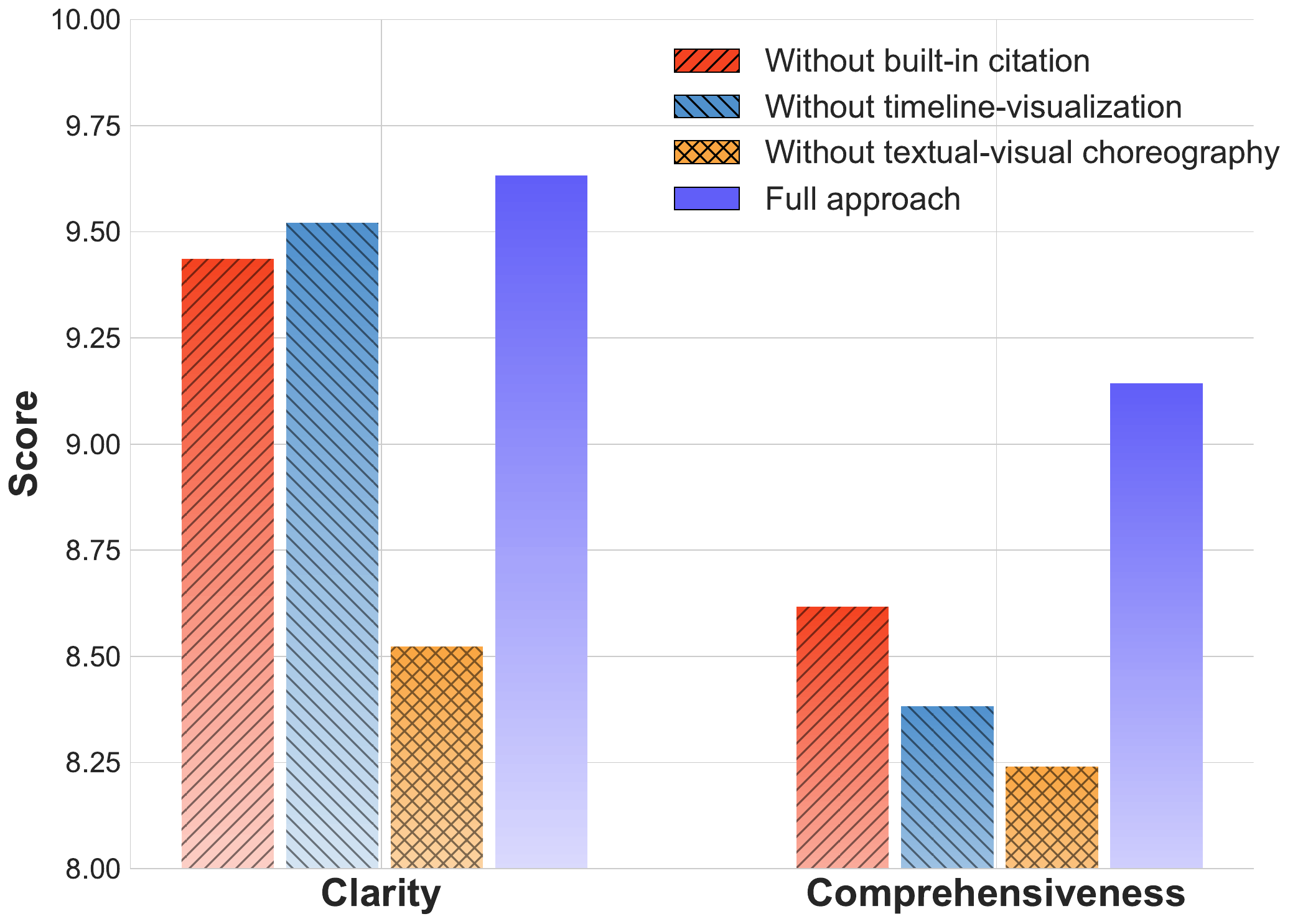}
  \caption{Ablation study of sub-modules for the rich answer representation.}
  \label{fig:ablation_module}
\end{figure}
% \subsubsection{Sub-module}
\paragraph{Query and Retrieved Documents Processing}
We first conduct an ablation study on the sub-modules designed to enhance text quality. Specifically, we compare the performance of the generated answers after omitting each sub-module against the full approach. LLM evaluation is used to rate the responses based on the multi-faceted evaluation criteria, with results provided in \cref{tab:ablation_study}.
Notably, skipping a sub-module does not always lead to a decline in all metrics. For example, omitting query expansion may improve relevance. However, the full approach, which integrates all sub-modules, achieves the best overall performance, as indicated by the average scores.

\paragraph{Representation Enhancement}
We further conduct an ablation study on built-in citation, timeline visualization, and textual-visual choreography to assess their impact on clarity and comprehensiveness based on human evaluation. As shown in \cref{fig:ablation_module}, removing any of these modules significantly reduces the comprehensiveness of the generated answer. Additionally, textual-visual choreography has a strong positive effect on clarity. These findings highlight the advantages of rich answer representations in supporting cognitive scaffolding and enhancing information assimilation efficiency.

\paragraph{Fine-Tuning}
In \ourmethod{}, we fine-tune multiple LLMs for generative tasks, including entity extraction, QDG generation, and response generation. To evaluate the impact of fine-tuning, we replace these models with GPT-4O and Qwen 2.5-72B. As shown in \cref{tab:ablation_model}, fine-tuned models enable \ourmethod{} to generate higher-quality answers.
Additionally, we present an ablation study on the fine-tuned generation model in \cref{tab:ablation_generation} and the reranking models in \cref{tab:ablation_rerank}.

\section{Conclusion}
\label{sec:conclusion}
In this work, we present \ourmethod{}, a generative AI search engine designed to tackle multi-faceted challenges in answer generation and user experience through a fully integrated pipeline. Our approaches not only build on state-of-the-art researches but also introduce novel solutions to specific challenges. Extensive experiments demonstrate the superiority of \ourmethod{} over existing technologies. Future work will focus on enhancing multilingual capabilities and expanding domain-specific optimizations.

\bibliographystyle{ACM-Reference-Format}
\bibliography{sample-base}

%%
%% If your work has an appendix, this is the place to put it.
\appendix
\section{Multi-Faceted Evaluation Criteria}
\label{app:eval}
The detailed definitions of the multi-faceted evaluation criteria are provided in \cref{tab:criteria}.
\section{Instruction Prompt}
\begin{table}[t]
    \captionsetup{justification=centering} % Centers the caption
    \caption{Multi-faceted evaluation criteria.}
    \label{tab:criteria}
    % \vspace{5pt} % Adds a little space between the caption and the table
    \centering
\begin{lstlisting}[language=Evaluation]
(1) Conciseness:
    - The response should directly address the user's question.
    - Avoid irrelevant content, unnecessary information, or roundabout explanations.
    - Deduct 1 point for each irrelevant statement.

(2) Numerical Precision:
    - If a question requires a specific number, avoid vague terms like "several" or "many times."
    - Responses should be precise and specific.
    - Deduct 1 point for each ambiguous statement.

(3) Relevance:
    - If the question specifies constraints (e.g., time, location, person, event), 
      the answer must adhere to them.
    - Deduct 1 point for each instance of misalignment with the question's constraints.

(4) Factuality:
    - The information must be factually correct, especially for numerical or factual questions.
    - Deduct 1 point for each incorrect numerical or factual statement.

(5) Timeliness:
    - For ongoing news or urgent reports, ensure information reflects the latest updates.
    - The current date is {to be filled}.
    - If the question is not time-sensitive, no points are deducted.
    - For time-sensitive questions, deduct points proportionally based on outdatedness.

(6) Comprehensiveness:
    - The response should comprehensively cover all aspects of the user's inquiry.
    - The user should not need further search to grasp the full context.
    - Deduct 1 point for each missing essential element.

(7) Clarity:
    - The response should be easy to understand, well-structured, and formatted logically.
    - Example: Chronological events should be presented in chronological order.
    - Deduct 1 point for unclear or disorganized presentation.

(8) Coherence:
    - The response should be logically consistent, with smooth transitions between sentences.
    - Deduct 1 point for each instance of incoherent or disjointed phrasing.

(9) Insightfulness:
    - The response should provide insightful or unique perspectives.
    - Base score: 6 points.
    - Award 0.5-1 additional points for each innovative idea or expression.
\end{lstlisting}
\end{table}

\subsection{Query-Decomposition Graph}
\label{app:qdt}
Prompt for the query-decomposition graph is provided in \cref{tab:query_analysis}.

\begin{table*}[t]
    \captionsetup{justification=centering} % Centers the caption
    \caption{Prompt for query-decomposition graph.}
    \label{tab:query_analysis}
    \centering
\begin{lstlisting}[language=Evaluation]
Please analyze the following query and return the 
explanation in dictionary format.

Response format:
{'is_complex': True/False, 'sub_queries': [], 'parent_child': []}

Analysis Steps and Principles:

1. **Classify the nature of the query**  
   - The query can be classified into one of two types:  
     (a) A "complex query" that consists of multiple sub-queries.  
     (b) A "simple query" that can be directly answered.  
   - If the query is classified as "complex," set 'is_complex' to **True**.  
   - If the query is "simple," set 'is_complex' to **False**, and leave 'sub_queries' and 'parent_child' as empty lists.

2. **Decomposing a Complex Query**  
   - If the query is classified as "complex," break it down into **sub-queries** and store them in the 'sub_queries' list.  
   - Decomposition principles:  
     1) If a query contains multiple **target entities**, split it into multiple sub-queries.  
        - Example: *"What are the latest social news and weather news in Shanghai?"*  
          - Target entities: *"social news"*, *"weather news"*.  
          - Split into: *"What are the latest social news in Shanghai?"* and *"What are the latest weather news in Shanghai?"*.  
     2) Each sub-query should be **indivisible** and should not require further decomposition.  
     3) No duplicate sub-queries.  
     4) When referring to **names of people, places, or organizations**, ensure full and precise descriptions.  
        - Example: *"What is the area and population of New Jersey, USA?"*  
          - Correct split: *"What is the area of New Jersey, USA?"* and *"What is the population of New Jersey, USA?"*.  
          - Incorrect split: *"What is the area of New Jersey?"* and *"What is the population of New Jersey?"*.  
     5) The total number of sub-queries **should not exceed 6**.

3. **Analyzing Dependencies Between Sub-Queries**  
   - If the query is complex, analyze the **dependency relationships** between sub-queries and store them in 'parent_child'.  
   - Example:  
     - *"What natural disasters occurred in Indonesia in April?"*  
     - *"How long did this natural disaster last?"*  
     - The second question **depends** on the first; thus, the first is the *parent*, and the second is the *child*:  
       ```json
       {"parent": "What natural disasters occurred in Indonesia in April?",  
        "child": "How long did this natural disaster last?"}
       ```
   - Dependency principles:  
     1) If sub-queries are **independent**, 'parent_child' remains an empty list.  
     2) If the **child question cannot be answered without the parent**, it is a dependent relationship.  
        - Example: "What is the latest iPhone model" is the parent node of "What are the specifications of the latest iPhone?"  
        - The first question must be answered before the second.  
     3) Every possible pair of sub-queries should be evaluated for dependency.  
        - A query can be both a *parent* and a *child* in different relationships.

### Example:
{Few-Shot Examples}

Query: {Query}  
Response: \n
\end{lstlisting}
\end{table*}
\subsection{Answer Generation}
\label{app:prompt:generation}
Prompt for answer generation is provided in  \cref{tab:encyclopedia_qa}.

\begin{table}[t]
    \captionsetup{justification=centering} % Centers the caption
    \caption{Prompt for answer generation.}
    \label{tab:encyclopedia_qa}
    \centering
\begin{lstlisting}[language=Evaluation]
You are an AI assistant named Xinyu, developed by the 
Shanghai Algorithm Innovation Research Institute. You are 
performing an encyclopedia Q\&A task. Please generate an 
answer based on the provided reference materials and 
related Q\&A content.

Question: {Sub-Query}

Related Q\&A:
{Ancestor Node 1: Sub-Query}
{Ancestor Node 1: Answer}
{Ancestor Node 2: Sub-Query}
{Ancestor Node 2: Answer}
...

Reference materials:
{Retrieved Passage 1}
{Retrieved Passage 2}
...

When generating your answer, follow these guidelines:

[Structural Requirements]:  
To ensure clarity and organization, you may use one or 
more of the following structured formats:
 - **Introduction-Body-Summary**: Introduce the topic, elaborate, and summarize key points.
 - **Paragraphs by Subquestion**: Address each subquestion in a separate paragraph.
 - **Cause and Effect**: Explain the causes and consequences of an event.
 - **Comparison and Contrast**: Describe and compare two or more concepts.
 - **Chronological Order**: Describe events or steps in order of occurrence.
 - **Problem-Solution**: Introduce a problem and explain solutions or strategies.
 - **Pros and Cons**: List the positive and negative aspects of a decision or choice.
 - **Definition and Examples**: Provide a definition and illustrate it with examples.
 - **Logical Reasoning**: Derive conclusions based on assumptions or premises.
 - **List Structure**: Enumerate facts or features for easy readability.
 - **Categorization**: Introduce a concept, group it by categories, and explain in detail.
 - **Theme and Variations**: Explore a core theme and its variations.
 - **Case Study**: Explain a theory or concept through specific cases.
 - **Hierarchical Structure**: Arrange information by importance or sequence.
 - **Issue and Counterarguments**: Present an issue with supporting and opposing views.

[Language Requirements]:  
(1) Use concise and clear language.  
(2) Ensure that the answer's structure enhances clarity and readability.  
(3) The response must directly and accurately address the question, avoiding irrelevant content.  
(4) When citing reference materials, ignore template formatting or improper phrasing.  
(5) If detailed elaboration is required, output the answer in a structured **Markdown** format.  

Your Answer: \n
\end{lstlisting}
\end{table}
\subsection{LLM Evaluation}
\label{app:ft:eval}
\Cref{tab:evaluation_prompt} presents the prompt used to instruct the LLM to evaluate the generated text based on multi-faceted criteria (see \cref{tab:criteria}). To reduce task complexity and enhance evaluation quality, we assess one facet per evaluation and fill the metric title and definition accordingly.
\begin{table}[t]
    \captionsetup{justification=centering} % Centers the caption
    \caption{Prompt for multi-faceted evaluation.}
    \label{tab:evaluation_prompt}
    \centering
\begin{lstlisting}[language=Evaluation]
Assume you are an article quality inspector. Please 
evaluate the response based on {Metric Title}. 
I will provide the user's question and the final response 
The maximum score is 10 points, and the scoring rules are 
as follows:

{Metric Definition}

Please strictly follow the scoring rules. Example output 
format: 

'{
    "Issues Identified": "X", 
    "Calculation Process": "10-1.0-1.0-1.0 = 7.0", 
    "Score": 7
}'

{Few-Shot Examples}

Your final score: \n"
\end{lstlisting}
\end{table}

\subsection{Built-In Citation}
\label{app:promp:cite}
Prompt for entity extraction is provided in \cref{tab:info_extraction}. Prompt for citation identification is provided in \cref{tab:citation_source_matching}.

\begin{table}[t]
    \captionsetup{justification=centering} % Centers the caption
    \caption{Prompt for entity extraction.}
    \label{tab:info_extraction}
    \centering
\begin{lstlisting}[language=Evaluation]
Read the given sentence and extract the contained 
information about time, location, persons, and job titles. 

Your extraction result should be returned in JSON format, 
with each field name restricted to one of the following: 
["Time", "Location", "Persons", "Job Titles"]

If there are multiple pieces of information of the same 
type in the sentence, the corresponding category's value 
should be represented as an array. 

Below are some examples:

{Few-Shot Examples}

{Sentence}

Extraction result: \n
\end{lstlisting}
\end{table}

\begin{table}[t]
    \captionsetup{justification=centering} % Centers the caption
    \caption{Prompt for citation identification.}
    \label{tab:citation_source_matching}
    \centering
\begin{lstlisting}[language=Evaluation]
You are a journalist skilled in analyzing the correlation 
between document information. I will provide you with a 
sentence excerpted from a news article, along with 
several reference documents used in writing this article. 
Your task is to determine which reference document the 
excerpted sentence most likely originates from.

The excerpted sentence is:
{Sentence}

The key information contained in this sentence is:
Time: {Time}
Location: {Location}
Person: {Person}
Job Title: {Job Title}
Numbers: {Numbers}

The reference documents used for writing this article and 
their respective key information are as follows:
[1] {Retrieved Document}
[2] {Retrieved Document}
[3] ...

When making your determination, ensure that the selected 
reference document matches as much key information from 
the excerpted sentence as possible. The higher the degree 
of key information overlap, the more likely the reference 
document is the source of the excerpted sentence.

Your response should contain only a one- or two-digit 
number representing the corresponding reference document 
number, such as "[2]", "[9]", or "[13]". If you believe 
that none of the reference documents are relevant to the 
given sentence, return "-1".

The most likely source document number is: \n
\end{lstlisting}
\end{table}
\section{Retrieval Documents Filtering Rules}
\label{app:filter_rule}
\paragraph{HTML Content Filtering} Non-content HTML tags, such as <script> and <style>, are removed using the lxml library to parse the original HTML into a DOM tree. Using XPath selectors, we retain only essential content tags like <div>, <p>, and <article>, ensuring compatibility with irregular HTML structures.

\paragraph{Text Processing}: Extracted text blocks are separated by spaces or line breaks to improve readability. Redundant whitespace and excessive line breaks are removed while preserving paragraph structure. Distracting elements like “Read More,” “Click to Continue,” or inline emojis are filtered out. Special characters, stopwords (e.g., from publicly available resources like Stopwords JSON\footnote{https://github.com/6/stopwords-json/blob/master/dist/ca.json}), emoji patterns (via regular expressions targeting Unicode Emoji ranges), and irrelevant newline characters are eliminated.
\paragraph{Sensitive Information Filtering}  Personal identifiers, such as phone numbers, email addresses, and platform-specific markers are detected and removed.
\paragraph{Text Normalization}
Punctuation is standardized to half-width characters, and numbers in various formats are converted to standardized half-width Arabic numerals.

\section{Fine-Tuning}
\label{app:fine-tune}
\subsection{User Intent Understanding}
\label{app:ft:intent}
\begin{figure}
\centering
\includegraphics[width=\linewidth]{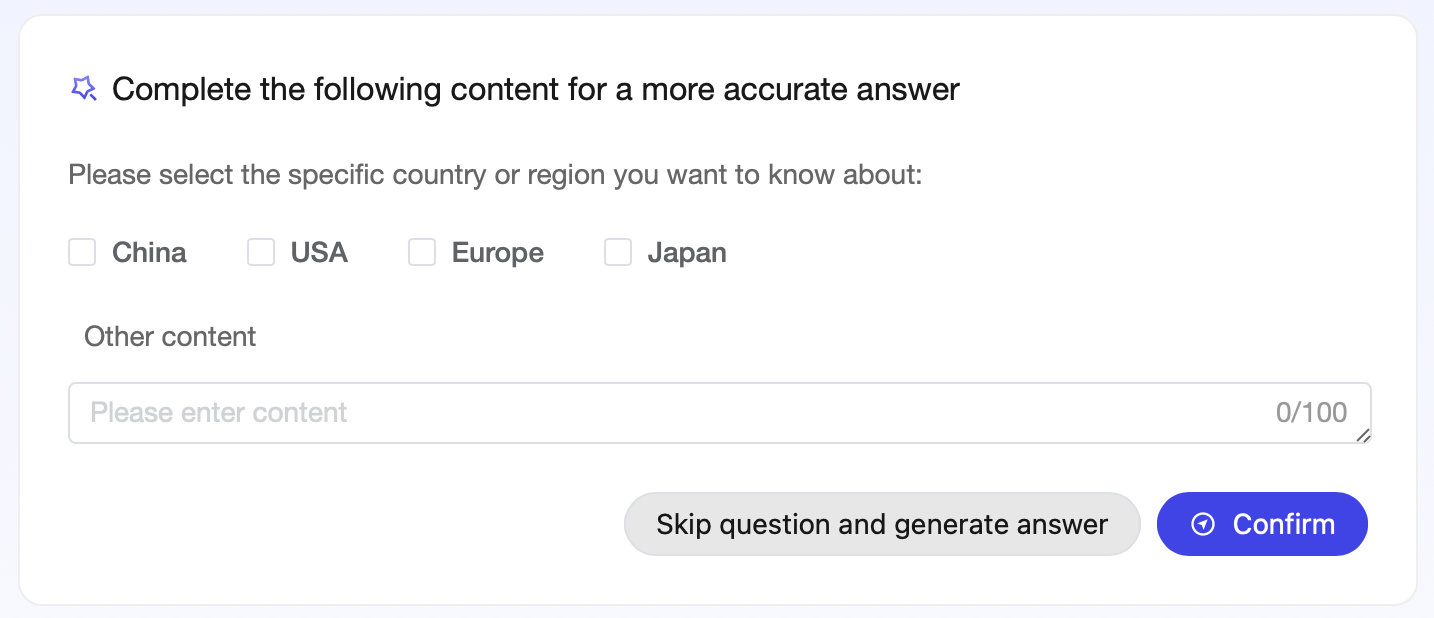}
  \caption{\ourmethod{}'s Interface for query disambiguation.}
  \label{fig:disambiguation}
\end{figure}
To promote responsible AI behavior and constrain the scope of queries, we fine-tune the model for the query rejection module. We define 11 categories of queries that warrant refusal, including: (1) illegal content, (2) ethical violations, (3) privacy breaches, (4) harmful intent, (5) professional consultations, (6) human-AI interactions, (7) misinformation, (8) technical inquiries, (9) academic requests, (10) planning and consulting inquiries, and (11) creative content generation.

To construct a training dataset, we collect a set of seed queries based on open-source datasets: Do-Not-Answer~\cite{wang2023donotanswer}, BeaverTails~\cite{ji2024beavertails} and Safety-Prompts~\cite{sun2023safety}. Additionally, we generate synthetic queries to compensate the imbalance number for each class in the collected dataset. Then we instruct multiple LLMs by given the definition of the 11 categories to classify and output in JSON format as follows:
\begin{verbatim}
{
“Refusal”: “Yes/No”,
“Category”: “illegal content/ethical viloations/…”
}
\end{verbatim}
Decisions are aggregated using a majority voting mechanism to establish a consensus. Finally, human experts review the results to correct misclassification.

For queries classified as non-refusal, we further prepare a dataset for query disambiguation. We instruct multiple LLMs to analyze whether a query requires further clarification to generate an appropriate response, outputting the results in the following JSON format:
\begin{verbatim}
{
  "Requires additional input": "Yes/No",
  "Additional options": {
    "Prompt description": "Please select...",
    "Choices": ["xx", "xx", ...]
  }
}
\end{verbatim}
Additional options are prepared to present clarifying options for the user as shown in \cref{fig:disambiguation}. Again, we use a majority voting mechanism to determine whether additional input is required. This process results in 7K data points. Human experts then review the results, refine the choices, and select high-quality samples. Ultimately, we construct a dataset of 5K data points, where 1/4 of the queries require additional input.

Two models are fine-tuned for the query rejection and query disambiguation tasks using their respective datasets, following \cref{eq:ntp}.

\subsection{Question-Decomposition Graph}
\label{app:ft:qdg}
We collect a set of queries and instruct multiple models using the prompt provided in \cref{app:qdt} to generate QDGs. The generated QDGs are then programmatically validated to ensure that the parent-child relationships meet the specified requirements, and duplicate QDGs are removed. This process results in 8,000 data points. Finally, human experts examine the data and select high-quality, correctly generated samples. In the end, we retain 4,184 data points for fine-tuning, which is conducted based on \cref{eq:ntp}.

\subsection{Reranker Model}
\label{app:fine-tune-rerank}
We construct a dataset comprising question-answer pairs using recent real-world data and public datasets such as MS MARCO~\cite{bajaj2016ms}. To generate hard negatives for fine-tuning, we apply various chunking strategies to create multiple candidate samples resembling the positive examples. These candidates are then ranked based on a base reranker model, selecting the top-300 samples. Next, we leverage multiple LLMs to assess whether each generated sample can answer the corresponding question. If the majority vote is negative, the sample is designated as a hard negative. In total, we generate 56K pairs. Finally, human experts review the results and curate a high-quality subset, retaining 13K pairs for fine-tuning, which is performed based on \cref{eq:reranker}.

\begin{table}[t!]
    \centering
    \caption{Ablation study of fine-tuning the rerank model.}
    \label{tab:ablation_rerank}
    \resizebox{\linewidth}{!}{ % Adjust to full width
    \begin{tabular}{l c c c c}
        \specialrule{1.2pt}{0pt}{0pt} % Thicker top rule
        \small \textbf{Model} & 
        \small \textbf{Precision} & 
        \small \textbf{Recall} & 
        \small \textbf{F1 Score} & 
        \small \textbf{Wall Time (s)} \\
        \midrule
        \small GPT-4O~\cite{chatgpt}             & 0.717 & 0.719 & {0.692} & 3.6  \\
        \small Qwen 2.5-72B~\cite{yang2024qwen2}     & 0.541 & {0.894} & 0.641 & 2.4 \\
        \small bge-reranker-v2-m3~\cite{chen2024bge}     & 0.568 & 0.671 & 0.562 & 0.1 \\
        \small Xinyu (ours)                              & 0.607 & 0.735 & 0.623 & 0.1  \\
        \specialrule{1.2pt}{0pt}{0pt} % Thicker bottom rule
    \end{tabular}
    } % Close resizebox
\end{table}
\begin{table*}[t!]
    \centering
    \caption{Multi-faceted comparison of different approaches based on GPT-4O (gpt-4-0125-preview). Higher value indicates better performance, 10 is the maximum.}
    \label{tab:performance_comparison_llm}
    \resizebox{\textwidth}{!}{
    \begin{tabular}{l c c c c c c c c c c}
        \specialrule{1.2pt}{0pt}{0pt}
        Model & \textbf{Conciseness} & \textbf{Numerical Precision} & \textbf{Relevance} & \textbf{Factuality} & \textbf{Timeliness} & \textbf{Comprehensiveness} & \textbf{Clarity} & \textbf{Coherence} & \textbf{Insightfulness} & \textbf{Average} \\
        \midrule
        Perplexity AI & \textbf{9.913} & \textbf{9.607} & \textbf{9.740} & 9.727 & 8.120 & 8.280 & 9.887 & \textbf{9.853} & 6.613 & 9.082 \\
        Tiangong AI~\cite{tiangong} & 9.819 & 9.188 & 9.570 & 9.738 & 7.758 & 7.517 & 9.839 & 9.799 & 6.161 & 8.821 \\
        Ernie Bot~\cite{yiyan} & 9.814 & 9.152 & 9.556 & 9.648 & 8.062 & 7.924 & 9.745 & 9.814 & 6.552 & 8.918 \\
        KIMI~\cite{kimi} & 9.695 & 9.359 & 9.576 & 9.675 & 8.059 & 8.305 & 9.686 & 9.720 & 6.432 & 8.945 \\
        Metaso~\cite{metaso} & 9.781 & 8.932 & 9.493 & 9.596 & 7.589 & 6.842 & 9.712 & 9.589 & 5.801 & 8.593 \\
        ChatGLM~\cite{chatglm} & 9.733 & 9.274 & 9.568 & \textbf{9.745} & 7.986 & 7.911 & 9.863 & 9.808 & 6.603 & 8.943 \\
        Baichuan~\cite{baichuan} & 9.433 & 9.053 & 9.307 & 9.403 & 7.813 & 7.832 & 9.373 & 9.200 & 6.640 & 8.673 \\
        Tongyi~\cite{tongyi} & 9.747 & 8.900 & 9.313 & 9.527 & 7.700 & 7.940 & 9.827 & 9.740 & 6.493 & 8.799 \\
       Xinyu (Ours) & 9.880 & 9.547 & 9.547 & 9.731 & \textbf{8.300} & \textbf{8.533} & \textbf{9.900} & 9.747 & \textbf{7.107} & \textbf{9.144 } \\
        \specialrule{1.2pt}{0pt}{0pt}
    \end{tabular}
    }
\end{table*}

\begin{table*}[t]
    \centering
    \caption{Ablation study of replacing our fine-tuned answer generation model with proprietary models.}
    \label{tab:ablation_generation}
    \resizebox{\linewidth}{!}{ % Make the table full width
    \begin{tabular}{l c c c c c c c c c c}
        \specialrule{1.2pt}{0pt}{0pt} % Thicker top rule
        \small \textbf{Model} & 
        \small \textbf{Conciseness} & 
        \small \textbf{Numerical Precision} & 
        \small \textbf{Relevance} & 
        \small \textbf{Factuality} & 
        \small \textbf{Timeliness} & 
        \small \textbf{Comprehensiveness} & 
        \small \textbf{Clarity} & 
        \small \textbf{Coherence} & 
        \small \textbf{Insightfulness} & 
        \small \textbf{Average} \\
        \midrule
        \small GPT-4O~\cite{chatgpt} & \textbf{9.854} & 9.482 & \textbf{9.588} & 9.597 & 8.107 & 8.515 & 9.849 & 9.734 & 6.989 & 9.080 \\
        \small Qwen 2.5-72B~\cite{yang2024qwen2} & 9.824 & 9.380 & 9.551 & 9.337 & 7.947 & 8.417 & 9.848 & 9.683 & 6.884 & 8.986 \\
        \small Xinyu (ours) & 9.880 & \textbf{9.547} & 9.547 & \textbf{9.731} & \textbf{8.300 }& \textbf{8.533} & \textbf{9.900} & \textbf{9.747} & \textbf{7.107} & \textbf{9.142} \\
        \specialrule{1.2pt}{0pt}{0pt} % Thicker bottom rule
    \end{tabular}
    } % Close resizebox
\end{table*}

\subsection{Generation}
\label{app:ft:generation}
We collect a set of queries and retrieve relevant documents, then generate responses using multiple LLMs and the prompt provided in \cref{app:prompt:generation}, yielding 121K answers. Human experts review the results, removing low-quality responses and refining the retained ones to ensure consistency in tone and eliminate hallucinations. Ultimately, 37K high-quality answers are selected for fine-tuning, which is performed using \cref{eq:ntp}.
\subsection{Built-In Citation}
\label{app:ft:cite}
As this module requires high efficiency, we fine-tune two SLMs (Qwen2.5-3B), using its larger counterpart, Qwen2.5-72B. We collect a set of passages and use Qwen2.5-72B to extract the entities from each sentence based on the prompt provided in \cref{app:promp:cite}. If any entities can be extracted, we retain the (passage, sentence, entities) triplet as part of the dataset. This process yields 33K data points. Human experts then review the data to correct errors and remove low-quality samples, ultimately retaining 26K data points.
We fine-tune the entity extraction SLM using sentences paired with their corresponding entities. Additionally, we fine-tune another SLM to retrieve the relevant passage from a given set based on the extracted entities, using the prompt provided in \cref{app:promp:cite}. Both models are optimized following \cref{eq:ntp}.
\section{Additional Results}
\label{app:res}

\paragraph{Multi-Faceted Evaluation by LLM}
\Cref{tab:performance_comparison_llm} presents the results of a multi-faceted evaluation conducted by GPT-4O. While the absolute values differ from those obtained through human evaluation (see \cref{tab:performance_comparison}), the rankings remain similar, demonstrating a strong correlation (see \cref{tab:pearson_correlation}). These results confirm that LLM-based evaluation is also indicative of performance.

\paragraph{Ablation of Fine-Tuning the Answer Generation Model}
\cref{tab:performance_comparison_llm} shows that replacing our fine-tuned answer generation model with either its base model (Qwen2.5-72B) or a strong proprietary model (GPT-4O) results in lower-quality generated answers. This finding indicates that our fine-tuning process is effective.

\paragraph{Ablation of Fine-Tuning the Reranker Model}
We compare our fine-tuned reranker model against its base model (bge-reranker-v2-m3) and two LLMs by instructing them to generate a ranking order based on relevance. As shown in \cref{tab:ablation_rerank}, our method preserves the efficiency of the base model while significantly outperforming it and achieving performance comparable to GPT-4o. The wall time for GPT-4o reflects API response time, whereas the wall time for other models is measured on a local cluster equipped with NVIDIA H800 GPUs. Batch size is set to 1.
\end{document}